\documentclass[aps,pra,twocolumn, superscriptaddress, showpacs, nofootinbib]{revtex4-1}
\usepackage{hyperref}
\usepackage{amsmath}
\usepackage{amssymb,amsfonts}
\usepackage{graphicx}
\usepackage{relsize}
\usepackage{lmodern}
\usepackage{scalefnt}
\usepackage{subfigure}
\usepackage{footnote}
\usepackage{xspace}
\usepackage{ifpdf}
\usepackage{pgffor}
\usepackage{array} 
\usepackage{stackrel}
\usepackage{ulem}
\newcolumntype{C}{>{$}c<{$}} 
\usepackage{tikz}
\usetikzlibrary{shapes.geometric}
\usepackage{color}
\usetikzlibrary{arrows,matrix,calc,scopes,decorations.markings}
\allowdisplaybreaks[1]

\makeatletter
\newcommand*{\Relbarfill@}{\arrowfill@\Relbar\Relbar\Relbar}
\newcommand*{\xeq}[2][]{\ext@arrow 0055\Relbarfill@{#1}{#2}}
\makeatother
\newcommand{\be}{ \begin{equation}}
\newcommand{\ee}{\end{equation}}

\def\bea{\begin{eqnarray}}
\def\eea{\end{eqnarray}}

\def\tr{\mathrm{tr}}

\begin{document}

 \title{Subregion duality, wedge classification and no global symmetries in AdS/CFT}

\author{Ning Bao}
 \email{ningbao75@gmail.com}
\affiliation{Department of Physics, Northeastern University, Boston, MA 02115, USA}
\affiliation{Computational Science Inititative, Brookhaven National Laboratory, Upton, NY 11973, USA}

\author{Yikun Jiang}
 \email{phys.yk.jiang@gmail.com}
\affiliation{Department of Physics, Northeastern University, Boston, MA 02115, USA}

\author{Joydeep Naskar}
 \email{naskar.j@northeastern.edu}
\affiliation{Department of Physics, Northeastern University, Boston, MA 02115, USA}
\affiliation{The NSF AI Institute for Artificial Intelligence and Fundamental Interactions, Cambridge, MA, U.S.A.}

\begin{abstract}
We study various notions of `subregion duality' in the context of AdS/CFT. We highlight the differences between the `background wedge' and the `operator reconstruction wedges,' providing a resolution to the paradox raised in \cite{Bao:2019hwq}. Additionally, we elucidate the distinctions between four different `operator reconstruction wedges' and demonstrate how to enhance the proof for the absence of global symmetries in geometrical states in AdS/CFT \cite{Harlow:2018jwu, Harlow:2018tng} as an example of these distinctions.
\end{abstract}
\maketitle

\section{Introduction}

The AdS/CFT correspondence \cite{Maldacena:1997re, Witten:1998qj} is a duality that relates a $d$-dimensional non-gravitational conformal field theory (CFT) to $(d+1)$-dimensional asymptotically anti-de Sitter (AdS) quantum gravity. This duality provides a non-perturbative definition of quantum gravity, prompting the question of how the degrees of freedom in the CFT are mapped to those in one higher dimension. Specifically, we seek to understand if the mapping is sufficiently local to associate a subset of the bulk that's `exactly dual' to a reduced density matrix $\rho_A$ for a boundary subregion $A$. This question, first raised in \cite{Czech:2012bh, Bousso:2012sj, Bousso:2012mh} is termed `subregion duality.' As a `duality,' the answer to this question would provide something in the bulk that contains exactly the same information as the boundary reduced density matrix $\rho_A$. Major advancements on this question come from the study of entanglement entropy, particularly the `quantum extremal surface' (QES) formula \eqref{QES},
and its connection to quantum error correction from a series of works \cite{Ryu:2006bv, Hubeny:2007xt, Faulkner:2013ana, Engelhardt:2014gca, Jafferis:2015del, Almheiri:2014lwa, Dong:2016eik, Faulkner:2017vdd} that we will review in \ref{operator wedge}. The upshot of the discovery is that there is an `entanglement wedge' (EW) in the bulk for the boundary subregion $A$. Using the boundary reduced density matrix $\rho_A$, we can reconstruct everything in EW$(A)$ from $A$, but nothing in its complement. This `entanglement wedge reconstruction' (EWR) thus provides an answer to the question of `subregion duality.' Moreover, assuming EWR, it was proven in \cite{Harlow:2018tng, Harlow:2018jwu} that there is no global symmetry in quantum gravity in the context of geometrical states in AdS/CFT.

However, this is not the end of the story. In \cite{Akers:2020pmf, Akers:2023fqr}, it was shown that the QES formula needs to be corrected even at leading order in large $N$ or $1/G_N$ expansion. Therefore, we cannot use the proposed procedure from \cite{Jafferis:2015del, Almheiri:2014lwa, Dong:2016eik, Faulkner:2017vdd} to reconstruct everything in EW$(A)$, which questions the validity of the `duality.' In fact, it was proposed in \cite{Akers:2020pmf, Akers:2023fqr} that the problem of reconstructing the bulk subregion from the boundary is not directly related to entanglement entropies, but is actually a problem of `one-shot state merging.' Using ideas from `one-shot state merging,' it was proposed that there is a wedge $R(A)$ that's in general smaller than EW$(A)$ where we can reconstruct all operators, and another larger wedge $G(A)$ beyond which we cannot reconstruct any operators. On the other hand, it was proven in \cite{Bao:2019bib, Bao:2019hwq, Bao:2020abm} that there is a systematic procedure for determining the bulk metric from the boundary data, leading to a bulk wedge $B(A)$ that's larger than EW$(A)$, providing another challenge for the duality. In this note, we want to clarify that there is no single bulk subregion that is exactly `dual to' and contains exactly the same information as boundary reduced density matrix $\rho_A$. Instead, for different aspects and physical questions, we have different `subregion dualities,' and we will explain five different bulk wedges that are related to $\rho_A$ and their associated `subregion dualities.'
\begin{figure}
	\centering
	\includegraphics[width=0.5\linewidth]{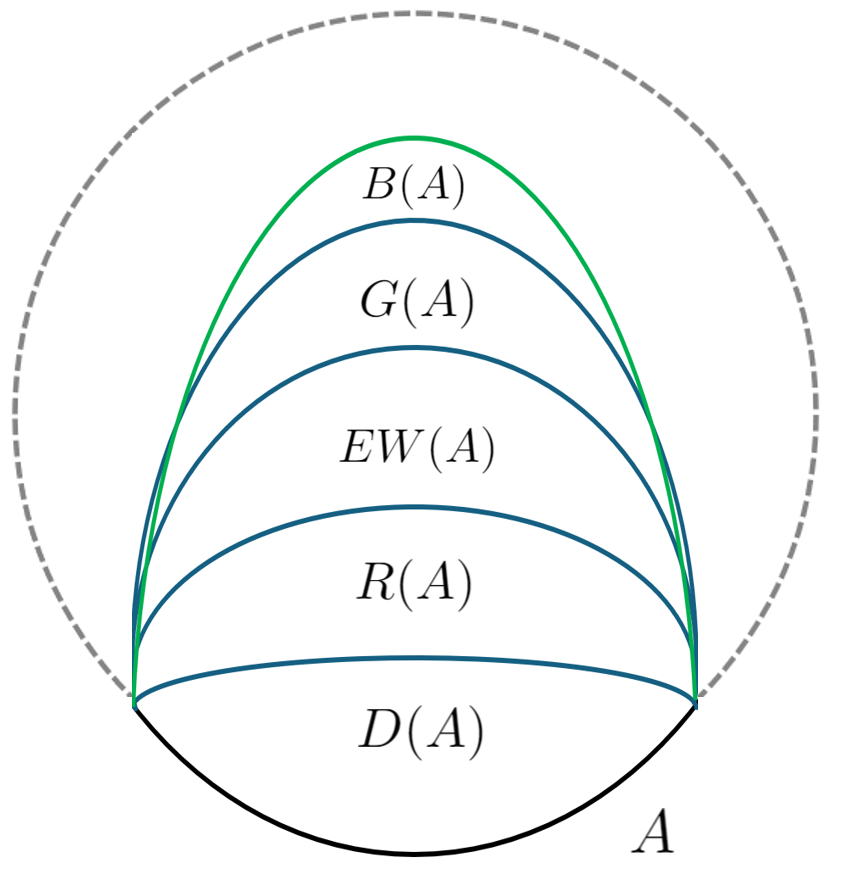}
	\caption{Five different wedges corresponding to reduced density matrix $\rho_A$ on subregion $A$. The wedge in green B(A) is the one associated to background reconstruction, and the wedges in blue are related to perturbative operator reconstuction on a fixed background. }
	\label{fivewedges}
\end{figure}

We want to first emphasize that the problem of reconstructing the bulk in the large $N$ limit from a boundary reduced density matrix really contains two steps, due to the special feature of a theory of quantum gravity, as described below. First, we need to make sure that the density matrix is `geometrical' in the large $N$ limit, i.e., it has a well-defined weakly coupled semi-classical geometrical description. This is implicitly assumed in many discussions in the literature, but we must keep in mind that not all states in the CFT have this property. For these states, we need to determine the bulk metric and treat it as a background before discussing reconstructing operators on top of this background. This step is often implicit but is essential, as in quantum gravity we are dealing with dynamical gravity that universally couples to everything, and its Hilbert space is generally not well-understood. It is in the large $N$ limit, where the gravitational interactions are largely suppressed, that we understand the gravitational Hilbert space to be approximately separated into fluctuations around different fixed background sectors.\footnote{At finite $N$, the quantum gravity Hilbert space and the notion of `bulk subregions' are not well-understood.} The Hilbert space and algebraic structure of quantum gravity in the large $N$ limit have recently been analyzed in great detail in \cite{Witten:2021jzq, Leutheusser:2021frk, Leutheusser:2022bgi}. Only after we have determined the gravitational background sector we want to study can the usual notion of `bulk operator reconstruction'  be defined. In short, we need to build up the spacetimes\cite{VanRaamsdonk:2010pw} before we can build up small perturbations on top of the spacetimes.

With this two-step procedure in mind, we can discuss different wedges for `subregion dualities'. This note is divided into two main parts. In Section \ref{bgd}, we first explain the definition and properties of the `background wedge' $B(A)$ and point out why the observation made in \cite{Bao:2019hwq} regarding a larger wedge than the `operator reconstruction wedges' is not problematic. Then, in Section \ref{operator wedge}, we illustrate four different bulk `operator reconstruction wedges.' After clarifying their distinctions, we highlight an improvement required in the assumptions made in the proof for the absence of global symmetries in geometrical states in AdS/CFT \cite{Harlow:2018jwu, Harlow:2018tng}. Using this example, we show how to upgrade the arguments using the appropriate bulk operator reconstruction wedge, thereby elucidating the different physical meanings between these wedges.

\section{Five different wedges}
\subsection{Background wedge}\label{bgd}

The first wedge is the wedge where we can reconstruct the metric from the reduced density matrix $\rho_A$, which we call the `background wedge' $B(A)$. The problem of extracting information for the background bulk metric from the boundary data was initiated in \cite{Balasubramanian:1999zv}. The idea is to translate `non-local' correlations on the boundary CFT into geometrical quantities in the bulk to learn the background geometry. It was proposed to look at the boundary two point correlation functions, which can be written in terms of the boundary reduced density matrix as
\be
G(x,y)=\tr_A (\rho_A O(x) O(y)) ,
\ee
where $x,y$ are the coordinates on the boundary CFT, and $O$ is the boundary CFT operator dual to bulk heavy scalar field $\phi$, with mass $1/l_{\text{AdS}} \ll m \ll 1/l_{\text{Planck}},1/l_{\text{string}}$. Via the AdS/CFT correspondence, $G(x,y)$ is equal to the boundary limit of two point function for the dual bulk fields. The mass of the field is in the regime where we can calculate this quantity using the geodesic approximation (WKB approximation) and neglect backreactions. Concretely, we have
\be
G(x,y)\approx e^{-m L_{ren}}
\ee
where $L_{ren}$ is the renormalized geodesic length through the bulk, connecting the two points on the boundary.

Recovering the unique bulk metric from the lengths of boundary-anchored geodesics is the `boundary rigidity problem'\cite{Porrati:2003na}, and this problem currently does not have a definitive answer in asymptoticaly AdS spacetimes. However, it has been proven in \cite{Bao:2019bib, Bao:2020abm} that we can uniquely reconstruct the bulk metric by considering a similar problem, using the areas of boundary anchored two dimensional extremal surfaces.

In boundary CFTs, the quantity related to these two-dimensional areas is the boundary Wilson loop expectation value. By introducing a probe D-brane and taking the decoupling limit\cite{Maldacena:1998im, Rey:1998ik, Drukker:1999zq}, we can relate the Wilson loops on the boundary CFTs to two-dimensional extremal surfaces (worldsheets) in the AdS bulk. Similar to boundary two-point functions, the saddle point approximation relates the Wilson loop expectation value to the exponential of the area for the two dimensional extremal surfaces. Thus using the theorem shown in \cite{Bao:2019bib, Bao:2020abm}, we can reconstruct the metric in the union of these two-dimensional boundary-anchored extremal surfaces in the bulk. Of course, as emphasized in \cite{Louko:2000tp, Giddings:2001pt}, the saddle point approximation has certain built-in assumptions and does not always relate these boundary correlation functions to bulk geometrical quantities, especially in timelike situations. A sufficient condition for the validity of these approximations is complex analyticity of the bulk metric, but this is not a necessary condition. We leave a better understanding for the exact condition under which these approximations apply for the future. There is another type of quantity that is related to bulk geometrical objects, namely the correspondence between entanglement entropies and bulk codimensional-two extremal surfaces\cite{Ryu:2006bv, Hubeny:2007xt, Faulkner:2013ana, Engelhardt:2014gca}\footnote{In the special case of three/four bulk dimensions, this reduces to the geodesic lengths/2D extremal surface areas we considered above, and has been studied from this perspective in \cite{Czech:2015qta,  Cao:2020uvb, Bao:2019bib}.}. We can obtain these quantities by further tracing out degrees of freedom in $\rho_A$, and calculate the von-Neumann entropy. Then using the correspondence, we can read off the  boundary-anchored codimensional-two extremal areas.
\begin{figure}
	\centering
	\includegraphics[width=1\linewidth]{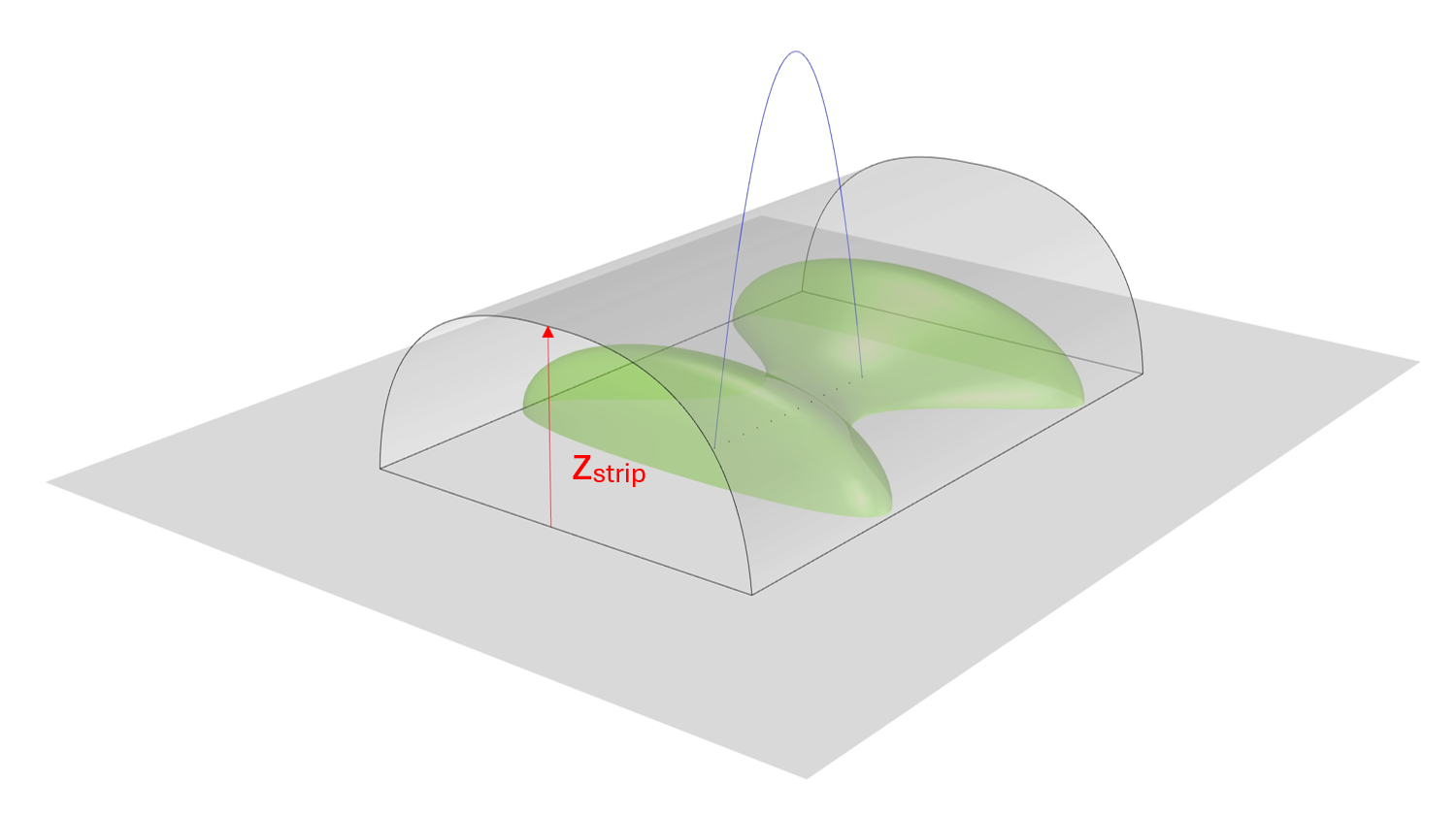}
	\caption{The blue curve is an extremal surface with a codimension higher than two, and extends deeper into the bulk than the codimensional two extremal surface, which is shaded green. The depth of the green surface is bounded by the depth $z_{strip}$ for the extremal surface of the strip.}
	\label{zstrip}
\end{figure}
We do not want to exhaust all the possible boundary objects giving bulk geometrical quantities, but we want to emphasize that, in principle, there should be a background wedge, which is distinct from the operator reconstruction wedges to be discussed in the later sections. In \cite{Bao:2019hwq}, it was found that there are situations where we can identify a much larger background wedge than the `entanglement wedge' that we will discuss later. This is due to the fact that the union of extremal surfaces with higher codimensions can probe deeper into the bulk than codimension-two extremal surfaces in those situations. See Fig. \ref{zstrip} for an illustration. The blue curve represents an extremal surface with a codimension higher than two, and extends deeper than the codimensional two extremal surface which is shaded green.  We want to point out that this problem of getting a larger wedge than the entanglement wedge persists as long as there is a way to extract the low-dimensional extremal surface areas/lengths. Even if the geodesic/Wilson loop approximation is not applicable, completely ruling out a way to extract them from anything calculable from the boundary reduced density matrix $\rho_A$ seems too strong a claim without proof. The point we want to make in this paper is that this is not surprising, as there is not a single subregion duality, and these wedges have \textit{completely different} physical meanings. In fact, since operator reconstruction does not even make sense without a fixed background, it is probably even natural to expect a larger background wedge in generic cases. We will also see an example where this naive expectation fails.

\subsection{Operator reconstruction wedges}\label{operator wedge}
The problem of bulk subregion operator reconstruction is the following: `Having fixed the whole bulk background metric, what part of the operators in the bulk can be expressed in terms of those on the boundary subregion?' We will call the wedges related to this question as `operator reconstruction wedges.' As was originally pointed out in \cite{Czech:2012bh}, it is natural to expect that there shall be bulk regions $G(A)$ where some operators can be reconstructed, and nothing in the bulk beyond $G(A)$ is contained in the boundary subregion. On the other hand, there shall also be $R(A)$, where all operators can be reconstructed. Before we discuss these wedges, we first explain the causal wedge
$D(A)$, which is immediately determined after we fix the bulk geometry.

\subsubsection{Causal wedge $D(A)$}
The causal wedge $D(A)$ is defined as the intersection of the causal past and causal future for the domain of dependence $D_A$ of $A$ (so it is the region in the bulk that $D_A$ can communicate with)\cite{Czech:2012bh}. Since we have a semi-classical bulk description in the large $N$ limit, we can study quantum field theories on curved spacetimes \cite{Hamilton:2005ju, Hamilton:2006az}. Using equations of motion on a fixed background, we can solve and express the fields in the bulk causal wedge $D(A)$ from boundary operators in $D_A$. This means $D(A) \subseteq  R(A)$. 
\newline
\newline
\newline

However, $D(A)$ is generally smaller than $R(A)$. The reason we can get larger wedges than $D(A)$ is that we can take advantage of quantum entanglement for the reconstruction. More explicitly, the entanglement entropies, or the von-Neumann entropies for the boundary and bulk reduced density matrices $S=- \tr (\rho \ln\rho)$, are proposed to be related by the remarkable quantum extremal surface (QES) formula\cite{Ryu:2006bv, Hubeny:2007xt, Faulkner:2013ana, Engelhardt:2014gca}
\be \label{QES}
S_{A}=\text{min ext}_{\gamma}\left(\frac{\text{Area}(\gamma)}{4G_N}+S_{\text{bulk}}(C)\right)
\ee
where $C$ is a bulk surface with A being part of its boundary, i.e., $\partial C=\gamma \cup A$, and $S_{\text{bulk}}(C)$ is the entanglement entropy in the bulk low energy effective field theory. The minimal extremal surface is the `quantum extremal surface,' and will be denoted as $\gamma_A$. The domain of dependence for the corresponding codimenional-one surface $C(A)$ will be called the `entanglement wedge' $\text{EW}(A)$.

Based on this formula, JLMS \cite{Jafferis:2015del} showed the perturbative equivalence at leading order in $N$ between boundary and bulk modular Hamiltonians,
\be \label{JLMS}
-\ln \rho_A=K_A=K_{C(A)}+\frac{\hat{\text{Area}}(\gamma_A)}{4 G_N}
\ee
We can define the modular flow of operators using these modular Hamiltonians. \cite{Almheiri:2014lwa, Dong:2016eik, Faulkner:2017vdd} further argued that \eqref{JLMS} is true to all orders in perturbation of $N$, and used ideas from quantum error correction to show the `Entanglement Wedge Reconstruction' in the `code subspace,' namely states where the backreaction to the metric is perturbatively small. This essentially says
\be \label{EWR WEDGES}
G(A)=\text{EW}(A)=R(A)
\ee

Many interesting and important results for our understanding of quantum gravity start from this result, including a proposal for the calculation of the Page curve in the context of AdS/CFT coupled to a bath\cite{Almheiri:2019psf, Penington:2019npb}. However, it was found in \cite{Akers:2020pmf} that the QES formula must be corrected at leading order in $N$, and is only true for `perfectly compressible states' for the matter fields (for example, the pure or thermal states that was mostly considered in literature before). Then since the QES formula \eqref{QES} is not valid for all the quantum states, \eqref{EWR WEDGES} is not true anymore.

Does this make it impossible to find the wedges $R(A)$ and $G(A)$? The modern point of view proposed in \cite{Akers:2020pmf, Akers:2023fqr} is that the question of operator reconstruction does not rely on the specific entanglement measure of entanglement entropy. Instead, the problem of encoding bulk subregions into boundary subregions is really a question about flow of quantum information for the specific state $\rho_A$, and thus manifestly a question of `one-shot state merging.' It is more convenient to phrase the problem in the Schroedinger picture instead of the Heisenberg picture above. In the Schroedinger picture, this problem can be briefly summarized as follows: suppose Alice and Bob share a mixed state $\rho_{AB}$, how many qubits does Alice need to send to Bob such that almost all the information is held by Bob? In quantum information theory, it is shown that the minimal number of qubits required for this task is given by the smooth conditional max-entropy $H_{\text{max}}^\epsilon(AB|B)$. It was thus proposed in \cite{Akers:2020pmf} that the correct `state-specific' duals for $R(A)$ and $G(A)$ are the `max-entanglement wedge' (max-EW) and the `min-entanglement wedge' (min-EW). To simplify the discussion, we will focus on spacetimes with a reflection symmetric time slice; the conjectural covariant generalizations can be found in \cite{Akers:2023fqr}.

\subsubsection{R(A) as \text{Max-EW}}

To understand the proposed $R(A)$ for which we can reconstruct everything from the boundary subregion $A$, we first define the smooth conditional max-entropy $H_{\text{max}}^\epsilon(AB|B)_\rho$ as
\be
H_{\text{max}}^\epsilon(AB|B)_{\rho}=\inf_{\tilde{\rho} \in \mathcal{B}^\epsilon(\rho)} \sup_{\sigma_B} \ln\left(\tr_A\left[\sqrt{\sigma_B^{1/2} \tilde{\rho}_{AB} \sigma_B^{1/2}}\right]\right)^2
\ee
\begin{figure}
	\centering
	\includegraphics[width=0.5\linewidth]{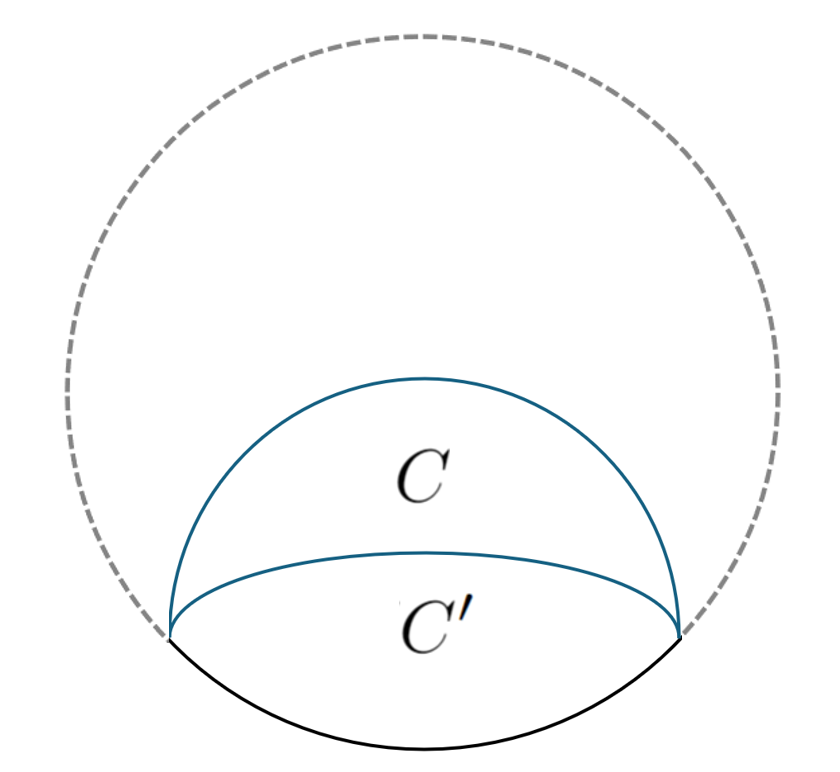}
	\caption{The max-EW C is defined such that all bulk subwedges satisfy \eqref{maxew}.}
	\label{maxwedge}
\end{figure}
where $\mathcal{B}^\epsilon(\rho)$ is the set of density matrices whose purified distance\footnote{The definition and properties of the purified distance can be found in \cite{Akers:2020pmf, Akers:2023fqr}. However, as emphasized in \cite{Akers:2020pmf}, the choice of the distance measure will only affect an unimportant scaling of $\epsilon$.} with $\rho_{AB}$ is smaller than a small number $\epsilon$. This $\epsilon$ provides the `smoothing,' manifesting the approximate nature of bulk operator reconstruction. This quantity intuitively quantifies how far our density of state $\rho_{AB}$ is from a decoupled state $\mathbf{1}_A \otimes \sigma_B$. The max-EW is then defined as the largest region C such that:
\be \label{maxew}
\forall C' \subset C, H_{\text{max}}^\epsilon(C|C')_{\rho_{\text{matter}}} < \frac{\text{Area}(\gamma_{C'})-\text{Area}(\gamma_C)}{4 G_N} ,
\ee
as shown in Fig. \ref{maxwedge}. Here we are writing it following \cite{Akers:2020pmf} by explicitly comparing the matter entanglement and background areas (which specifies the background entanglement), a more covariant proposal can be found in \cite{Akers:2023fqr}. It is shown in \cite{Akers:2020pmf, Akers:2023fqr} that such a largest region exists and satisfies properties such as nesting that we expect for operator reconstruction wedges. Moreover, it is argued that this is the bulk subregion where everything can be reconstructed from $A$, so it shall be identified with $R(A)$.

\subsubsection{G(A) as \text{Min-EW}}

The Min-EW is defined using the smooth conditional min-entropy $H_{\text{min}}^\epsilon(AB|B)_\rho$ as
\be
H_{\text{min}}^\epsilon(AB|B)_\rho=\sup_{\tilde{\rho} \in \mathcal{B}^\epsilon(\rho)} \left(-\min_{\sigma_B} \inf \{ \lambda: \tilde{\rho}_{AB} \leq e^{\lambda} \mathbf{1}_A \otimes \sigma_B\} \right)
\ee

The smooth conditional min-entropy and max-entropy satisfy the duality property: for any $\psi \in \mathcal{H}_A \otimes \mathcal{H}_B \otimes \mathcal{H}_C$,
\be \label{duality}
H_{\text{min}}^\epsilon(AB|B)_\psi=-H_{\text{max}}^\epsilon(AC|C)_\psi
\ee

The min-EW is defined as the smallest region C such that,
\be \label{minew}
\forall \bar{C}' \subset \bar{C}, H_{\text{min}}^\epsilon(C \bar{C}'|C)_{\rho_{\text{matter}}} > \frac{\text{Area}(\gamma_{C})-\text{Area}(\gamma_{C \bar{C}'})}{4 G_N}
\ee
where $\bar{C}'$ denotes the compliment region of $C'$. For pure states on the boundary, this is the complement region for the max wedge of $\bar{A}$. In general, the physical meaning of the min wedge is that no information outside this bulk region can flow to the boundary subregion, so it shall be identified as the bulk wedge $G(A)$.

\subsubsection{\text{EW(A)}}

Now we see that actually $R(A)$ $\neq$ $G(A)$ from one-shot state-merging, we can ask their connection to the originally proposed EW($A$) defined by \eqref{QES}. In fact, the smooth max/min-entropy encodes the size of the smallest/largest eigenvalues in the density matrix. Therefore, in general, EW($A$) would lie between $G(A)$ and $R(A)$ as shown in Fig.\ref{fivewedges}, also see explicit examples in \cite{Akers:2020pmf,Akers:2023fqr} where the three wedges are different. In a matter state with(approximately) flat spectrum like the pure or thermal states, the smooth max/min-entropy will unsurprisingly become equal to the von-Neumann entropy which captures the averaged eigenvalue, and in this case the three wedges would coincide, i.e, $R(A)$=EW($A$)=$G(A)$.

When we have a large number $n$ copies of the state, the conditional von-Neumann entropy is the asymptotic limit of the conditional smooth max and min entropies, 
\be
\begin{aligned}
\lim_{n \to \infty} \frac{1}{n} H_{\text{max}}^\epsilon(A^n|B^n)_{\psi^{\otimes n}}&=S(A|B)_\psi \\ &=\lim_{n \to \infty} \frac{1}{n} H_{\text{min}}^\epsilon(A^n|B^n)_{\psi^{\otimes n}}
\end{aligned}
\ee
However, as emphasized in \cite{Akers:2020pmf, Akers:2023fqr}, in holography we only have one copy of the holographic state, so it shall always be the one-shot quantities instead of the asymptotic ones that really matter.

\subsection{Wedges in the Python's Lunch geometry}

\begin{figure}
	\centering
	\includegraphics[width=0.6\linewidth]{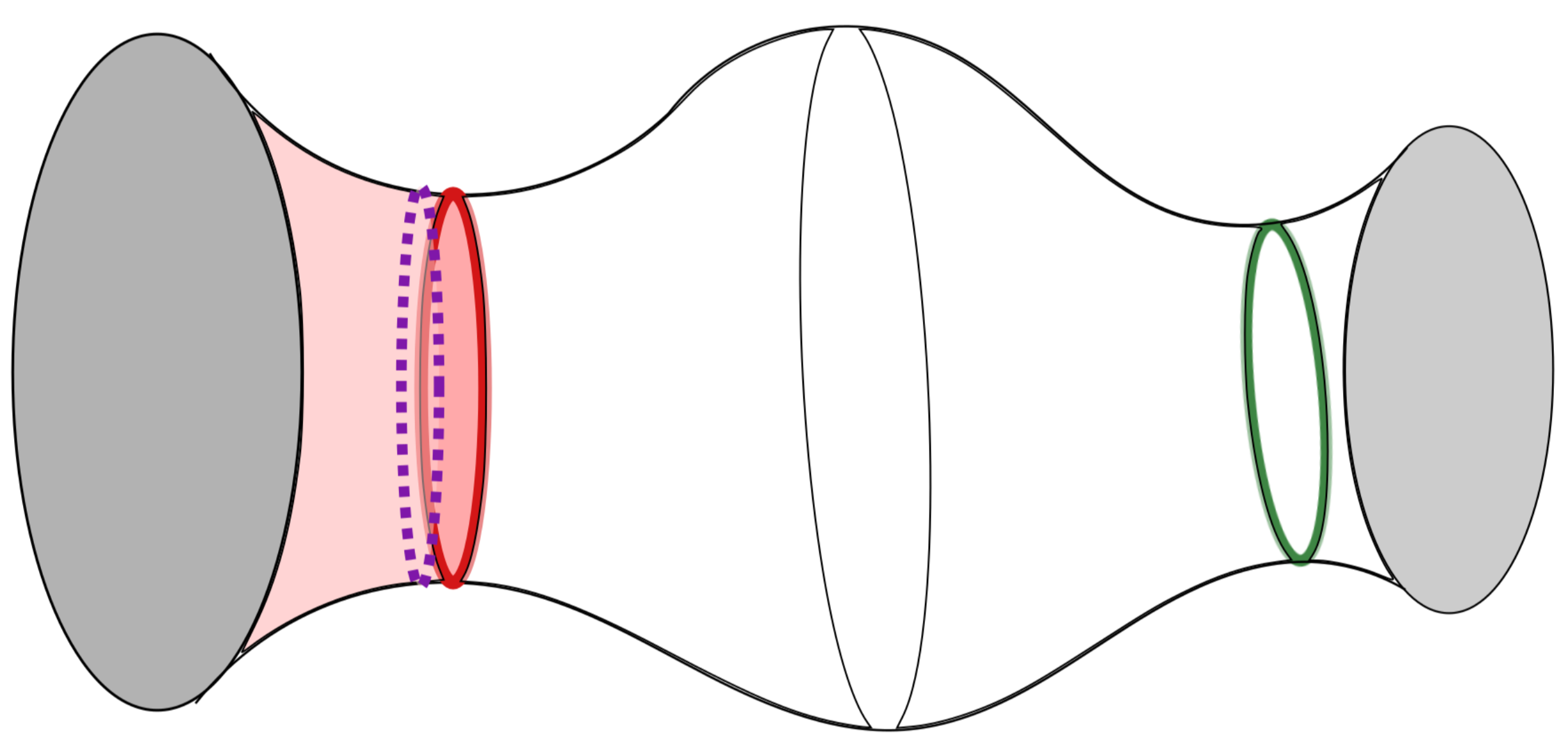}
	\caption{The Python’s Lunch wormhole geometry\cite{Brown:2019rox} features two minimal surfaces. In AdS$_3$\cite{Bao:2020hsc}, geodesics anchored to the left boundary cannot extend beyond the left bottleneck. As a result, the background wedge for the left asymptotic region is confined to the area between the boundary and the purple circle. For specific choices of matter entanglement, the Max-EW corresponds to the region enclosed by the left bottleneck, shown in red, while the Min-EW corresponds to the region bounded by the right bottleneck, shown in green.}
	\label{python}
\end{figure}

To further elucidate the distinctions between these wedges, we use the `Python’s Lunch' geometry \cite{Brown:2019rox} as an illustrative example. The Python’s Lunch is a wormhole geometry characterized by a bulge in the middle, as depicted in Fig.~\ref{python}. In this geometry, two local minimal surfaces serve as bottlenecks. We begin with the case of asymptotically AdS$_3$, where an explicit metric construction for the Python’s Lunch geometry is provided in \cite{Bao:2020hsc}. Let the lengths of the two bottlenecks be denoted by $\mathcal{A}_L$ and $\mathcal{A}_R$, with $\mathcal{A}_L > \mathcal{A}_R$ assumed without loss of generality.

Our focus will be on the distinct wedges within this two-dimensional time slice for the left asymptotic boundary. In this simplified setting, the only non-trivial geometrical objects are the geodesics. As noted in \cite{Bao:2020hsc}, the presence of a local minimum in the bulk homologous to the entire boundary ensures that no geodesic anchored at the boundary can extend beyond this minimal surface. Consequently, the background wedge is constrained to lie entirely between the boundary and the left bottleneck. In higher dimensions, geodesics or two-dimensional minimal surfaces \cite{Bao:2019bib, Bao:2020abm} similarly fail to provide a complete foliation of the entire region between the boundary and the right extremal surface. This distinction makes them clearly different from the Min-EW that we will explain in detail below.

Next, we determine the Max-EW and Min-EW. With two extremal surfaces homologous to the left boundary, the setup closely resembles that of \cite{Akers:2020pmf}. As argued in \cite{Akers:2020pmf}, for generic matter content with moderate bulk entropy gradients, it suffices to determine the Max-EW and Min-EW by examining regions perturbatively close to the two classical extremal surfaces. The two wedges depends on the exact matter entanglement in the theory for the criteria \eqref{maxew} and \eqref{minew}. Following the construction in \cite{Akers:2020pmf}, we can arrange scenarios that the Max-EW corresponds to the region bounded by the left extremal surface, while the Min-EW corresponds to the region bounded by the right extremal surface. It is noteworthy that, counterintuitively, the operator reconstruction wedges are generally larger than the background wedge in this case.

\section{No global symmetry in AdS/CFT}

Given that now we understand there are several different wedges in the bulk that are dual to a boundary subregion for different purposes, we can ask which ones are actually relevant for specific physical questions. Particularly, there were certain questions answered assuming EWR, or $R(A)$=EW($A$)=$G(A)$. We will use the proof for the absence of global symmetries for geometrical states in AdS/CFT \cite{Harlow:2018tng, Harlow:2018jwu} as an example,\footnote{The connection between the absence of global symmetry and subregion duality has also been studied in the context of generalized symmetries in \cite{Heckman:2024oot}.} and show how to upgrade their argument using the correct wedge that's actually needed.

\begin{figure}
	\centering
	\includegraphics[width=0.5\linewidth]{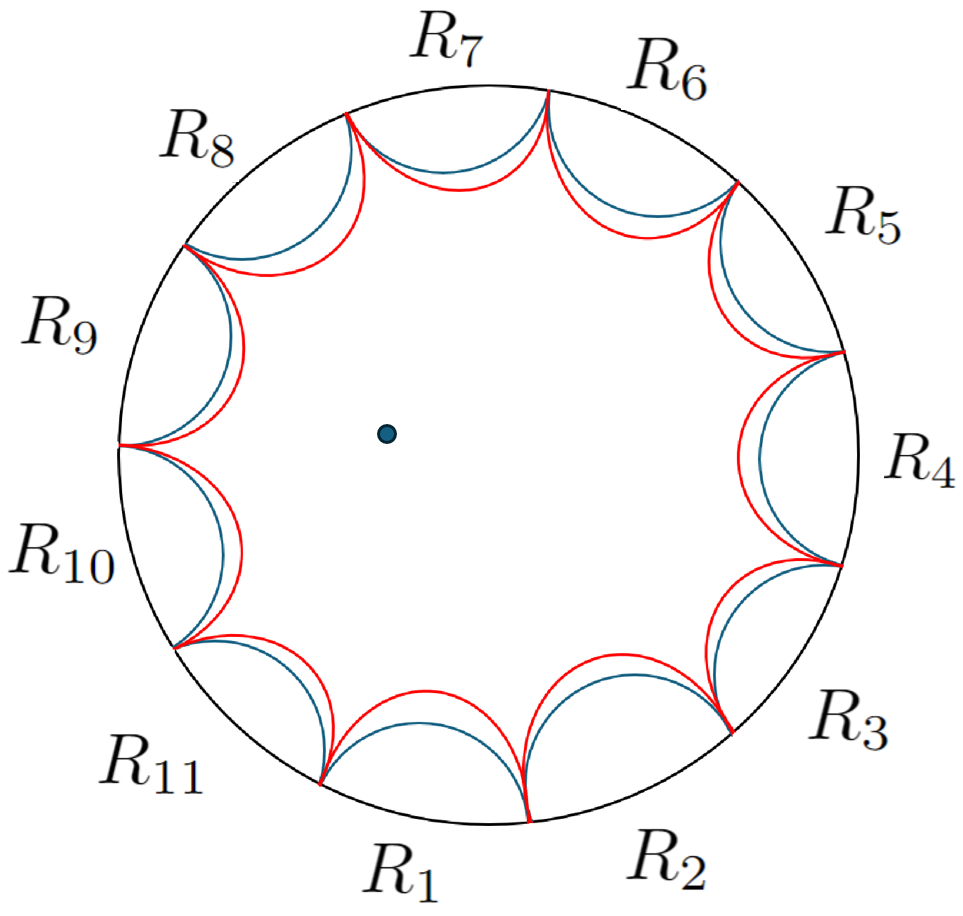}
	\caption{The CFT boundary is broken into disjoint pieces $R_i$, such that the union $\cup_i G(R_i)$ in the bulk bounded by the red curves  cannot reach the bulk point in the center, and thus the boundary global symmetry operators cannot act on operators defined on this point. The blue curves are the quantum extremal surfaces, which are contained in $\cup_i G(R_i)$.}
	\label{noglobalsymmetry}
\end{figure}
The argument in \cite{Harlow:2018tng, Harlow:2018jwu} goes as follows: 
\newline
\newline
1. Suppose there is global symmetry in the bulk quantum gravity, then by the extrapolating dictionary \cite{Witten:1998qj}, it gives a global symmetry on the boundary CFT.

2. We split the boundary Cauchy slice into a union of disjoint regions $R_i$. By the splittability of this boundary global symmetry, we have
\be \label{split}
U(g)=U(g,R_1)U(g,R_2)...U(g,R_n)U_{\text{edge}}
\ee
where $U_{\text{edge}}$ is defined at the boundaries of $R_i$ to fix the choices made in the definitions of $U(g,R_i)$. 

3. For any operators in the bulk, we can find small enough $R_i$, such that the union of their entanglement wedges $\cup_i \text{EW}(R_i)$ does not reach the local bulk operator. See Fig. \ref{noglobalsymmetry} where the blue curves are the QES for each $R_i$.

4. By EWR, the symmetry operators $U(g,R_i)$ can only affect bulk operators in $\text{EW}(R_i)$ but not its complement, so we know that any bulk operators cannot be charged under global symmetries, leading to a contradiction. 
\newline
\newline
We have seen above that EWR does not always hold as pointed out by \cite{Akers:2020pmf}, so the EWR assumption used in step 3 and step 4 of this proof does not apply. However, now that we understand that there is not a single duality for boundary subregions, we can ask if there is a way to upgrade this argument with the desired property wedge. Indeed, what really goes into the proof in step 3 and step 4 is the fact that due to certain locality properties in the boundary-bulk map, operators acting on certain boundary subregions can only affect a limited bulk subregion. We do not care if a bulk subregion is able to reconstruct everything as for $R(A)$, but we just need to make sure that nothing beyond the subregion can be reconstructed, and this is the definition of $G(A)$. Therefore, we can upgrade the proof in \cite{Harlow:2018tng, Harlow:2018jwu} by upgrading the usage of $\text{EW}(R_i)$ to $G(R_i)$, or $\text{min-EW}(R_i)$ according to \cite{Akers:2020pmf, Akers:2023fqr} as:
\newline
\newline
3’. For any operators in the bulk, we can find small enough $R_i$, such that the union of their min-wedges $\cup_i G(R_i)=\cup_i \text{min-EW}(R_i)$ does not reach the local bulk operator. 

\begin{figure}
	\centering
	\includegraphics[width=0.5\linewidth]{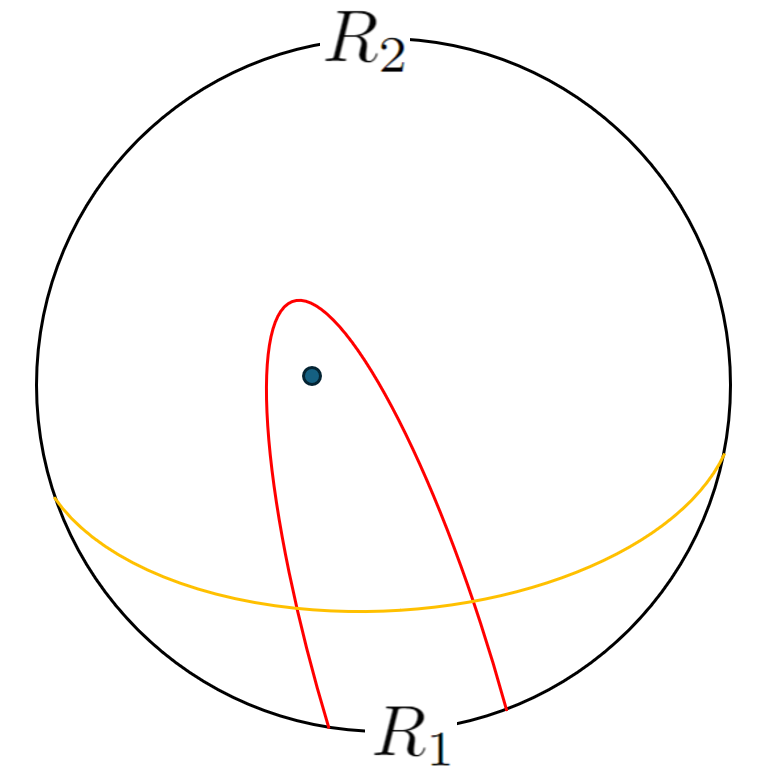}
	\caption{Situation where $G(R_1)$ goes very deep into the bulk and reaches the bulk operator can be ruled out by considering a large enough region in its complement and complementary recovery.}
	\label{GALARGE}
\end{figure}

See Fig. \ref{noglobalsymmetry} where $G(R_i)$ for each $R_i$ are bounded by the red curves. They contain a larger region in the bulk than $\text{EW}(R_i)$, which are bounded by the blue curves.

One might worry that there are some $G(R_i)$ that goes very deep in the bulk as in Fig. \ref{GALARGE}, such that it is impossible to choose $\cup_i G(R_i)$ not reaching certain bulk points. Here we will show that this cannot happen. Imagine we have such a small subregion $R_1$ where $G(R_1)$ goes very deep into the bulk and contains the bulk point. Then we can choose from the boundary complement $\bar{R}_1$ a large region $R_2$, such that $R(R_2)$ contains the bulk point. This is always possible if we look at a very tiny $R_1$ as $R(R_2)$ shall contain the causal wedge $D(R_2)$. Then the bulk operator will be reconstructible from both $R_1$ and $R_2 \subset \bar{R}_1$, violating complementary recovery. If we identify $G(A)$ as $\text{min-EW}(A)$, this is because of the duality property \eqref{duality}, and the $\text{min-EW}(A)$ is contained in the complement region of $\text{max-EW}(\bar{A})$.\footnote{These two are equal when the global state is pure. When the state is mixed, we can purify the state and see $\text{min-EW}(A) \subset \overline{\text{max-EW}(\bar{A})}$.} 

4'. By definition of $G(R_i)$, the symmetry operators $U(g,R_i)$ cannot affect any bulk operators in $\overline{G(R_i)}$. Thus the total symmetry operator \eqref{split} cannot affect any bulk operators, leading to a contradiction. 

To summarize, we see with this example, even without a single `subregion duality,' we can still rule out global symmetry for geometrical states in AdS/CFT.

\section{Discussions}

In this paper, we explore the concept of `subregion duality' in holography. We begin by emphasizing that bulk reconstruction for geometrical states is fundamentally a two-step process: first, reconstructing the background geometry from CFT data, and second, reconstructing low-energy operators on this fixed background. Correspondingly, this distinction gives rise to the `background wedge' and the `operator reconstruction wedges,' which diverge at the very first step. This provides a resolution to the puzzle raised in \cite{Bao:2019hwq}, where the background wedge extends beyond the conventionally discussed operator reconstruction wedges. Next, we clarify the roles of four distinct `operator reconstruction wedges,' highlighting that the one-shot versions of entanglement \cite{Akers:2020pmf, Akers:2023fqr} are central to addressing different problems in holography. To further illustrate these concepts, we use the Python's lunch geometry as a concrete example, showing the difference in these wedges. Finally, we refine the proof for the absence of global symmetries in geometrical states in AdS/CFT \cite{Harlow:2018jwu, Harlow:2018tng} by incorporating the proper operator reconstruction wedge, demonstrating the importance of distinguishing between these wedges in different physical contexts.

During the time this paper was under review, another work \cite{Akers:2024wre} appeared, also emphasizing on the distinction between `operator reconstruction' and `background reconstruction' (which they call `geometry reconstruction') from the perspective of complexity theory, where the authors argue that while operator reconstruction might be efficient, but geometry reconstruction can be generically hard. Another recent work \cite{Bhattacharjee:2024ceb} applied the concepts of min-EW and max-EW within the framework of uberholography \cite{Pastawski:2016qrs,Bao:2022tgv}, affirming that the phase transition between connected and disconnected phases on a fractal boundary geometry, is generically not an \textit{all or nothing} phenomenon.

\section*{Acknowledgements}
We thank Chris Akers, Gong Cheng, Wan Zhen Chua, Sebastian Fischetti,  Keiichiro Furuya, Hao Geng, Ling-Yan Hung, Hao Geng, Chen-Te Ma, Suvrat Raju, Grant Remmen, Martin Sasieta, Douglas Stanford, Aron Wall, Yixu Wang, Zixia Wei for helpful comments and discussions. We thank Zhenhao Zhou for plotting the 3D diagram for us. Y.J. acknowledges the support by Novel Quantum Algorithms from Fast Classical Transforms, the U.S Department of Energy ASCR EXPRESS grant, and Northeastern University. J.N. is partially supported by the NSF under Cooperative Agreement PHY2019786 and N.B.'s startup funding at Northeastern University. N.B. is supported by the Novel Quantum Algorithms from Fast Classical Transforms, the U.S Department of Energy ASCR EXPRESS grant, and Northeastern University.

\bibliography{main.bib}

\begin{thebibliography}{44}%
\makeatletter
\providecommand \@ifxundefined [1]{%
 \@ifx{#1\undefined}
}%
\providecommand \@ifnum [1]{%
 \ifnum #1\expandafter \@firstoftwo
 \else \expandafter \@secondoftwo
 \fi
}%
\providecommand \@ifx [1]{%
 \ifx #1\expandafter \@firstoftwo
 \else \expandafter \@secondoftwo
 \fi
}%
\providecommand \natexlab [1]{#1}%
\providecommand \enquote  [1]{``#1''}%
\providecommand \bibnamefont  [1]{#1}%
\providecommand \bibfnamefont [1]{#1}%
\providecommand \citenamefont [1]{#1}%
\providecommand \href@noop [0]{\@secondoftwo}%
\providecommand \href [0]{\begingroup \@sanitize@url \@href}%
\providecommand \@href[1]{\@@startlink{#1}\@@href}%
\providecommand \@@href[1]{\endgroup#1\@@endlink}%
\providecommand \@sanitize@url [0]{\catcode `\\12\catcode `\$12\catcode `\&12\catcode `\#12\catcode `\^12\catcode `\_12\catcode `\%12\relax}%
\providecommand \@@startlink[1]{}%
\providecommand \@@endlink[0]{}%
\providecommand \url  [0]{\begingroup\@sanitize@url \@url }%
\providecommand \@url [1]{\endgroup\@href {#1}{\urlprefix }}%
\providecommand \urlprefix  [0]{URL }%
\providecommand \Eprint [0]{\href }%
\providecommand \doibase [0]{http://dx.doi.org/}%
\providecommand \selectlanguage [0]{\@gobble}%
\providecommand \bibinfo  [0]{\@secondoftwo}%
\providecommand \bibfield  [0]{\@secondoftwo}%
\providecommand \translation [1]{[#1]}%
\providecommand \BibitemOpen [0]{}%
\providecommand \bibitemStop [0]{}%
\providecommand \bibitemNoStop [0]{.\EOS\space}%
\providecommand \EOS [0]{\spacefactor3000\relax}%
\providecommand \BibitemShut  [1]{\csname bibitem#1\endcsname}%
\let\auto@bib@innerbib\@empty
\bibitem [{\citenamefont {Bao}\ \emph {et~al.}(2020{\natexlab{a}})\citenamefont {Bao}, \citenamefont {Chatwin-Davies}, \citenamefont {Niehoff},\ and\ \citenamefont {Usatyuk}}]{Bao:2019hwq}%
  \BibitemOpen
  \bibfield  {author} {\bibinfo {author} {\bibfnamefont {N.}~\bibnamefont {Bao}}, \bibinfo {author} {\bibfnamefont {A.}~\bibnamefont {Chatwin-Davies}}, \bibinfo {author} {\bibfnamefont {B.~E.}\ \bibnamefont {Niehoff}}, \ and\ \bibinfo {author} {\bibfnamefont {M.}~\bibnamefont {Usatyuk}},\ }\href {\doibase 10.1103/PhysRevD.101.066011} {\bibfield  {journal} {\bibinfo  {journal} {Phys. Rev. D}\ }\textbf {\bibinfo {volume} {101}},\ \bibinfo {pages} {066011} (\bibinfo {year} {2020}{\natexlab{a}})},\ \Eprint {http://arxiv.org/abs/1911.00519} {arXiv:1911.00519 [hep-th]} \BibitemShut {NoStop}%
\bibitem [{\citenamefont {Harlow}\ and\ \citenamefont {Ooguri}(2019)}]{Harlow:2018jwu}%
  \BibitemOpen
  \bibfield  {author} {\bibinfo {author} {\bibfnamefont {D.}~\bibnamefont {Harlow}}\ and\ \bibinfo {author} {\bibfnamefont {H.}~\bibnamefont {Ooguri}},\ }\href {\doibase 10.1103/PhysRevLett.122.191601} {\bibfield  {journal} {\bibinfo  {journal} {Phys. Rev. Lett.}\ }\textbf {\bibinfo {volume} {122}},\ \bibinfo {pages} {191601} (\bibinfo {year} {2019})},\ \Eprint {http://arxiv.org/abs/1810.05337} {arXiv:1810.05337 [hep-th]} \BibitemShut {NoStop}%
\bibitem [{\citenamefont {Harlow}\ and\ \citenamefont {Ooguri}(2021)}]{Harlow:2018tng}%
  \BibitemOpen
  \bibfield  {author} {\bibinfo {author} {\bibfnamefont {D.}~\bibnamefont {Harlow}}\ and\ \bibinfo {author} {\bibfnamefont {H.}~\bibnamefont {Ooguri}},\ }\href {\doibase 10.1007/s00220-021-04040-y} {\bibfield  {journal} {\bibinfo  {journal} {Commun. Math. Phys.}\ }\textbf {\bibinfo {volume} {383}},\ \bibinfo {pages} {1669} (\bibinfo {year} {2021})},\ \Eprint {http://arxiv.org/abs/1810.05338} {arXiv:1810.05338 [hep-th]} \BibitemShut {NoStop}%
\bibitem [{\citenamefont {Maldacena}(1998{\natexlab{a}})}]{Maldacena:1997re}%
  \BibitemOpen
  \bibfield  {author} {\bibinfo {author} {\bibfnamefont {J.~M.}\ \bibnamefont {Maldacena}},\ }\href {\doibase 10.4310/ATMP.1998.v2.n2.a1} {\bibfield  {journal} {\bibinfo  {journal} {Adv. Theor. Math. Phys.}\ }\textbf {\bibinfo {volume} {2}},\ \bibinfo {pages} {231} (\bibinfo {year} {1998}{\natexlab{a}})},\ \Eprint {http://arxiv.org/abs/hep-th/9711200} {arXiv:hep-th/9711200} \BibitemShut {NoStop}%
\bibitem [{\citenamefont {Witten}(1998)}]{Witten:1998qj}%
  \BibitemOpen
  \bibfield  {author} {\bibinfo {author} {\bibfnamefont {E.}~\bibnamefont {Witten}},\ }\href {\doibase 10.4310/ATMP.1998.v2.n2.a2} {\bibfield  {journal} {\bibinfo  {journal} {Adv. Theor. Math. Phys.}\ }\textbf {\bibinfo {volume} {2}},\ \bibinfo {pages} {253} (\bibinfo {year} {1998})},\ \Eprint {http://arxiv.org/abs/hep-th/9802150} {arXiv:hep-th/9802150} \BibitemShut {NoStop}%
\bibitem [{\citenamefont {Czech}\ \emph {et~al.}(2012)\citenamefont {Czech}, \citenamefont {Karczmarek}, \citenamefont {Nogueira},\ and\ \citenamefont {Van~Raamsdonk}}]{Czech:2012bh}%
  \BibitemOpen
  \bibfield  {author} {\bibinfo {author} {\bibfnamefont {B.}~\bibnamefont {Czech}}, \bibinfo {author} {\bibfnamefont {J.~L.}\ \bibnamefont {Karczmarek}}, \bibinfo {author} {\bibfnamefont {F.}~\bibnamefont {Nogueira}}, \ and\ \bibinfo {author} {\bibfnamefont {M.}~\bibnamefont {Van~Raamsdonk}},\ }\href {\doibase 10.1088/0264-9381/29/15/155009} {\bibfield  {journal} {\bibinfo  {journal} {Class. Quant. Grav.}\ }\textbf {\bibinfo {volume} {29}},\ \bibinfo {pages} {155009} (\bibinfo {year} {2012})},\ \Eprint {http://arxiv.org/abs/1204.1330} {arXiv:1204.1330 [hep-th]} \BibitemShut {NoStop}%
\bibitem [{\citenamefont {Bousso}\ \emph {et~al.}(2012)\citenamefont {Bousso}, \citenamefont {Leichenauer},\ and\ \citenamefont {Rosenhaus}}]{Bousso:2012sj}%
  \BibitemOpen
  \bibfield  {author} {\bibinfo {author} {\bibfnamefont {R.}~\bibnamefont {Bousso}}, \bibinfo {author} {\bibfnamefont {S.}~\bibnamefont {Leichenauer}}, \ and\ \bibinfo {author} {\bibfnamefont {V.}~\bibnamefont {Rosenhaus}},\ }\href {\doibase 10.1103/PhysRevD.86.046009} {\bibfield  {journal} {\bibinfo  {journal} {Phys. Rev. D}\ }\textbf {\bibinfo {volume} {86}},\ \bibinfo {pages} {046009} (\bibinfo {year} {2012})},\ \Eprint {http://arxiv.org/abs/1203.6619} {arXiv:1203.6619 [hep-th]} \BibitemShut {NoStop}%
\bibitem [{\citenamefont {Bousso}\ \emph {et~al.}(2013)\citenamefont {Bousso}, \citenamefont {Freivogel}, \citenamefont {Leichenauer}, \citenamefont {Rosenhaus},\ and\ \citenamefont {Zukowski}}]{Bousso:2012mh}%
  \BibitemOpen
  \bibfield  {author} {\bibinfo {author} {\bibfnamefont {R.}~\bibnamefont {Bousso}}, \bibinfo {author} {\bibfnamefont {B.}~\bibnamefont {Freivogel}}, \bibinfo {author} {\bibfnamefont {S.}~\bibnamefont {Leichenauer}}, \bibinfo {author} {\bibfnamefont {V.}~\bibnamefont {Rosenhaus}}, \ and\ \bibinfo {author} {\bibfnamefont {C.}~\bibnamefont {Zukowski}},\ }\href {\doibase 10.1103/PhysRevD.88.064057} {\bibfield  {journal} {\bibinfo  {journal} {Phys. Rev. D}\ }\textbf {\bibinfo {volume} {88}},\ \bibinfo {pages} {064057} (\bibinfo {year} {2013})},\ \Eprint {http://arxiv.org/abs/1209.4641} {arXiv:1209.4641 [hep-th]} \BibitemShut {NoStop}%
\bibitem [{\citenamefont {Ryu}\ and\ \citenamefont {Takayanagi}(2006)}]{Ryu:2006bv}%
  \BibitemOpen
  \bibfield  {author} {\bibinfo {author} {\bibfnamefont {S.}~\bibnamefont {Ryu}}\ and\ \bibinfo {author} {\bibfnamefont {T.}~\bibnamefont {Takayanagi}},\ }\href {\doibase 10.1103/PhysRevLett.96.181602} {\bibfield  {journal} {\bibinfo  {journal} {Phys. Rev. Lett.}\ }\textbf {\bibinfo {volume} {96}},\ \bibinfo {pages} {181602} (\bibinfo {year} {2006})},\ \Eprint {http://arxiv.org/abs/hep-th/0603001} {arXiv:hep-th/0603001} \BibitemShut {NoStop}%
\bibitem [{\citenamefont {Hubeny}\ \emph {et~al.}(2007)\citenamefont {Hubeny}, \citenamefont {Rangamani},\ and\ \citenamefont {Takayanagi}}]{Hubeny:2007xt}%
  \BibitemOpen
  \bibfield  {author} {\bibinfo {author} {\bibfnamefont {V.~E.}\ \bibnamefont {Hubeny}}, \bibinfo {author} {\bibfnamefont {M.}~\bibnamefont {Rangamani}}, \ and\ \bibinfo {author} {\bibfnamefont {T.}~\bibnamefont {Takayanagi}},\ }\href {\doibase 10.1088/1126-6708/2007/07/062} {\bibfield  {journal} {\bibinfo  {journal} {JHEP}\ }\textbf {\bibinfo {volume} {07}},\ \bibinfo {pages} {062} (\bibinfo {year} {2007})},\ \Eprint {http://arxiv.org/abs/0705.0016} {arXiv:0705.0016 [hep-th]} \BibitemShut {NoStop}%
\bibitem [{\citenamefont {Faulkner}\ \emph {et~al.}(2013)\citenamefont {Faulkner}, \citenamefont {Lewkowycz},\ and\ \citenamefont {Maldacena}}]{Faulkner:2013ana}%
  \BibitemOpen
  \bibfield  {author} {\bibinfo {author} {\bibfnamefont {T.}~\bibnamefont {Faulkner}}, \bibinfo {author} {\bibfnamefont {A.}~\bibnamefont {Lewkowycz}}, \ and\ \bibinfo {author} {\bibfnamefont {J.}~\bibnamefont {Maldacena}},\ }\href {\doibase 10.1007/JHEP11(2013)074} {\bibfield  {journal} {\bibinfo  {journal} {JHEP}\ }\textbf {\bibinfo {volume} {11}},\ \bibinfo {pages} {074} (\bibinfo {year} {2013})},\ \Eprint {http://arxiv.org/abs/1307.2892} {arXiv:1307.2892 [hep-th]} \BibitemShut {NoStop}%
\bibitem [{\citenamefont {Engelhardt}\ and\ \citenamefont {Wall}(2015)}]{Engelhardt:2014gca}%
  \BibitemOpen
  \bibfield  {author} {\bibinfo {author} {\bibfnamefont {N.}~\bibnamefont {Engelhardt}}\ and\ \bibinfo {author} {\bibfnamefont {A.~C.}\ \bibnamefont {Wall}},\ }\href {\doibase 10.1007/JHEP01(2015)073} {\bibfield  {journal} {\bibinfo  {journal} {JHEP}\ }\textbf {\bibinfo {volume} {01}},\ \bibinfo {pages} {073} (\bibinfo {year} {2015})},\ \Eprint {http://arxiv.org/abs/1408.3203} {arXiv:1408.3203 [hep-th]} \BibitemShut {NoStop}%
\bibitem [{\citenamefont {Jafferis}\ \emph {et~al.}(2016)\citenamefont {Jafferis}, \citenamefont {Lewkowycz}, \citenamefont {Maldacena},\ and\ \citenamefont {Suh}}]{Jafferis:2015del}%
  \BibitemOpen
  \bibfield  {author} {\bibinfo {author} {\bibfnamefont {D.~L.}\ \bibnamefont {Jafferis}}, \bibinfo {author} {\bibfnamefont {A.}~\bibnamefont {Lewkowycz}}, \bibinfo {author} {\bibfnamefont {J.}~\bibnamefont {Maldacena}}, \ and\ \bibinfo {author} {\bibfnamefont {S.~J.}\ \bibnamefont {Suh}},\ }\href {\doibase 10.1007/JHEP06(2016)004} {\bibfield  {journal} {\bibinfo  {journal} {JHEP}\ }\textbf {\bibinfo {volume} {06}},\ \bibinfo {pages} {004} (\bibinfo {year} {2016})},\ \Eprint {http://arxiv.org/abs/1512.06431} {arXiv:1512.06431 [hep-th]} \BibitemShut {NoStop}%
\bibitem [{\citenamefont {Almheiri}\ \emph {et~al.}(2015)\citenamefont {Almheiri}, \citenamefont {Dong},\ and\ \citenamefont {Harlow}}]{Almheiri:2014lwa}%
  \BibitemOpen
  \bibfield  {author} {\bibinfo {author} {\bibfnamefont {A.}~\bibnamefont {Almheiri}}, \bibinfo {author} {\bibfnamefont {X.}~\bibnamefont {Dong}}, \ and\ \bibinfo {author} {\bibfnamefont {D.}~\bibnamefont {Harlow}},\ }\href {\doibase 10.1007/JHEP04(2015)163} {\bibfield  {journal} {\bibinfo  {journal} {JHEP}\ }\textbf {\bibinfo {volume} {04}},\ \bibinfo {pages} {163} (\bibinfo {year} {2015})},\ \Eprint {http://arxiv.org/abs/1411.7041} {arXiv:1411.7041 [hep-th]} \BibitemShut {NoStop}%
\bibitem [{\citenamefont {Dong}\ \emph {et~al.}(2016)\citenamefont {Dong}, \citenamefont {Harlow},\ and\ \citenamefont {Wall}}]{Dong:2016eik}%
  \BibitemOpen
  \bibfield  {author} {\bibinfo {author} {\bibfnamefont {X.}~\bibnamefont {Dong}}, \bibinfo {author} {\bibfnamefont {D.}~\bibnamefont {Harlow}}, \ and\ \bibinfo {author} {\bibfnamefont {A.~C.}\ \bibnamefont {Wall}},\ }\href {\doibase 10.1103/PhysRevLett.117.021601} {\bibfield  {journal} {\bibinfo  {journal} {Phys. Rev. Lett.}\ }\textbf {\bibinfo {volume} {117}},\ \bibinfo {pages} {021601} (\bibinfo {year} {2016})},\ \Eprint {http://arxiv.org/abs/1601.05416} {arXiv:1601.05416 [hep-th]} \BibitemShut {NoStop}%
\bibitem [{\citenamefont {Faulkner}\ and\ \citenamefont {Lewkowycz}(2017)}]{Faulkner:2017vdd}%
  \BibitemOpen
  \bibfield  {author} {\bibinfo {author} {\bibfnamefont {T.}~\bibnamefont {Faulkner}}\ and\ \bibinfo {author} {\bibfnamefont {A.}~\bibnamefont {Lewkowycz}},\ }\href {\doibase 10.1007/JHEP07(2017)151} {\bibfield  {journal} {\bibinfo  {journal} {JHEP}\ }\textbf {\bibinfo {volume} {07}},\ \bibinfo {pages} {151} (\bibinfo {year} {2017})},\ \Eprint {http://arxiv.org/abs/1704.05464} {arXiv:1704.05464 [hep-th]} \BibitemShut {NoStop}%
\bibitem [{\citenamefont {Akers}\ and\ \citenamefont {Penington}(2021)}]{Akers:2020pmf}%
  \BibitemOpen
  \bibfield  {author} {\bibinfo {author} {\bibfnamefont {C.}~\bibnamefont {Akers}}\ and\ \bibinfo {author} {\bibfnamefont {G.}~\bibnamefont {Penington}},\ }\href {\doibase 10.1007/JHEP04(2021)062} {\bibfield  {journal} {\bibinfo  {journal} {JHEP}\ }\textbf {\bibinfo {volume} {04}},\ \bibinfo {pages} {062} (\bibinfo {year} {2021})},\ \Eprint {http://arxiv.org/abs/2008.03319} {arXiv:2008.03319 [hep-th]} \BibitemShut {NoStop}%
\bibitem [{\citenamefont {Akers}\ \emph {et~al.}(2024{\natexlab{a}})\citenamefont {Akers}, \citenamefont {Levine}, \citenamefont {Penington},\ and\ \citenamefont {Wildenhain}}]{Akers:2023fqr}%
  \BibitemOpen
  \bibfield  {author} {\bibinfo {author} {\bibfnamefont {C.}~\bibnamefont {Akers}}, \bibinfo {author} {\bibfnamefont {A.}~\bibnamefont {Levine}}, \bibinfo {author} {\bibfnamefont {G.}~\bibnamefont {Penington}}, \ and\ \bibinfo {author} {\bibfnamefont {E.}~\bibnamefont {Wildenhain}},\ }\href {\doibase 10.21468/SciPostPhys.16.6.144} {\bibfield  {journal} {\bibinfo  {journal} {SciPost Phys.}\ }\textbf {\bibinfo {volume} {16}},\ \bibinfo {pages} {144} (\bibinfo {year} {2024}{\natexlab{a}})},\ \Eprint {http://arxiv.org/abs/2307.13032} {arXiv:2307.13032 [hep-th]} \BibitemShut {NoStop}%
\bibitem [{\citenamefont {Bao}\ \emph {et~al.}(2019)\citenamefont {Bao}, \citenamefont {Cao}, \citenamefont {Fischetti},\ and\ \citenamefont {Keeler}}]{Bao:2019bib}%
  \BibitemOpen
  \bibfield  {author} {\bibinfo {author} {\bibfnamefont {N.}~\bibnamefont {Bao}}, \bibinfo {author} {\bibfnamefont {C.}~\bibnamefont {Cao}}, \bibinfo {author} {\bibfnamefont {S.}~\bibnamefont {Fischetti}}, \ and\ \bibinfo {author} {\bibfnamefont {C.}~\bibnamefont {Keeler}},\ }\href {\doibase 10.1088/1361-6382/ab377f} {\bibfield  {journal} {\bibinfo  {journal} {Class. Quant. Grav.}\ }\textbf {\bibinfo {volume} {36}},\ \bibinfo {pages} {185002} (\bibinfo {year} {2019})},\ \Eprint {http://arxiv.org/abs/1904.04834} {arXiv:1904.04834 [hep-th]} \BibitemShut {NoStop}%
\bibitem [{\citenamefont {Bao}\ \emph {et~al.}(2021)\citenamefont {Bao}, \citenamefont {Cao}, \citenamefont {Fischetti}, \citenamefont {Pollack},\ and\ \citenamefont {Zhong}}]{Bao:2020abm}%
  \BibitemOpen
  \bibfield  {author} {\bibinfo {author} {\bibfnamefont {N.}~\bibnamefont {Bao}}, \bibinfo {author} {\bibfnamefont {C.}~\bibnamefont {Cao}}, \bibinfo {author} {\bibfnamefont {S.}~\bibnamefont {Fischetti}}, \bibinfo {author} {\bibfnamefont {J.}~\bibnamefont {Pollack}}, \ and\ \bibinfo {author} {\bibfnamefont {Y.}~\bibnamefont {Zhong}},\ }\href {\doibase 10.1088/1361-6382/abcfd0} {\bibfield  {journal} {\bibinfo  {journal} {Class. Quant. Grav.}\ }\textbf {\bibinfo {volume} {38}},\ \bibinfo {pages} {047001} (\bibinfo {year} {2021})},\ \Eprint {http://arxiv.org/abs/2009.07850} {arXiv:2009.07850 [hep-th]} \BibitemShut {NoStop}%
\bibitem [{\citenamefont {Witten}(2021)}]{Witten:2021jzq}%
  \BibitemOpen
  \bibfield  {author} {\bibinfo {author} {\bibfnamefont {E.}~\bibnamefont {Witten}},\ }\href@noop {} {\  (\bibinfo {year} {2021})},\ \Eprint {http://arxiv.org/abs/2112.11614} {arXiv:2112.11614 [hep-th]} \BibitemShut {NoStop}%
\bibitem [{\citenamefont {Leutheusser}(2023)}]{Leutheusser:2021frk}%
  \BibitemOpen
  \bibfield  {author} {\bibinfo {author} {\bibfnamefont {S.~A.~W.}\ \bibnamefont {Leutheusser}},\ }\href {\doibase 10.1103/PhysRevD.108.086020} {\bibfield  {journal} {\bibinfo  {journal} {Phys. Rev. D}\ }\textbf {\bibinfo {volume} {108}},\ \bibinfo {pages} {086020} (\bibinfo {year} {2023})},\ \Eprint {http://arxiv.org/abs/2112.12156} {arXiv:2112.12156 [hep-th]} \BibitemShut {NoStop}%
\bibitem [{\citenamefont {Leutheusser}\ and\ \citenamefont {Liu}(2022)}]{Leutheusser:2022bgi}%
  \BibitemOpen
  \bibfield  {author} {\bibinfo {author} {\bibfnamefont {S.}~\bibnamefont {Leutheusser}}\ and\ \bibinfo {author} {\bibfnamefont {H.}~\bibnamefont {Liu}},\ }\href@noop {} {\  (\bibinfo {year} {2022})},\ \Eprint {http://arxiv.org/abs/2212.13266} {arXiv:2212.13266 [hep-th]} \BibitemShut {NoStop}%
\bibitem [{\citenamefont {Van~Raamsdonk}(2010)}]{VanRaamsdonk:2010pw}%
  \BibitemOpen
  \bibfield  {author} {\bibinfo {author} {\bibfnamefont {M.}~\bibnamefont {Van~Raamsdonk}},\ }\href {\doibase 10.1142/S0218271810018529} {\bibfield  {journal} {\bibinfo  {journal} {Gen. Rel. Grav.}\ }\textbf {\bibinfo {volume} {42}},\ \bibinfo {pages} {2323} (\bibinfo {year} {2010})},\ \Eprint {http://arxiv.org/abs/1005.3035} {arXiv:1005.3035 [hep-th]} \BibitemShut {NoStop}%
\bibitem [{\citenamefont {Balasubramanian}\ and\ \citenamefont {Ross}(2000)}]{Balasubramanian:1999zv}%
  \BibitemOpen
  \bibfield  {author} {\bibinfo {author} {\bibfnamefont {V.}~\bibnamefont {Balasubramanian}}\ and\ \bibinfo {author} {\bibfnamefont {S.~F.}\ \bibnamefont {Ross}},\ }\href {\doibase 10.1103/PhysRevD.61.044007} {\bibfield  {journal} {\bibinfo  {journal} {Phys. Rev. D}\ }\textbf {\bibinfo {volume} {61}},\ \bibinfo {pages} {044007} (\bibinfo {year} {2000})},\ \Eprint {http://arxiv.org/abs/hep-th/9906226} {arXiv:hep-th/9906226} \BibitemShut {NoStop}%
\bibitem [{\citenamefont {Porrati}\ and\ \citenamefont {Rabadan}(2004)}]{Porrati:2003na}%
  \BibitemOpen
  \bibfield  {author} {\bibinfo {author} {\bibfnamefont {M.}~\bibnamefont {Porrati}}\ and\ \bibinfo {author} {\bibfnamefont {R.}~\bibnamefont {Rabadan}},\ }\href {\doibase 10.1088/1126-6708/2004/01/034} {\bibfield  {journal} {\bibinfo  {journal} {JHEP}\ }\textbf {\bibinfo {volume} {01}},\ \bibinfo {pages} {034} (\bibinfo {year} {2004})},\ \Eprint {http://arxiv.org/abs/hep-th/0312039} {arXiv:hep-th/0312039} \BibitemShut {NoStop}%
\bibitem [{\citenamefont {Maldacena}(1998{\natexlab{b}})}]{Maldacena:1998im}%
  \BibitemOpen
  \bibfield  {author} {\bibinfo {author} {\bibfnamefont {J.~M.}\ \bibnamefont {Maldacena}},\ }\href {\doibase 10.1103/PhysRevLett.80.4859} {\bibfield  {journal} {\bibinfo  {journal} {Phys. Rev. Lett.}\ }\textbf {\bibinfo {volume} {80}},\ \bibinfo {pages} {4859} (\bibinfo {year} {1998}{\natexlab{b}})},\ \Eprint {http://arxiv.org/abs/hep-th/9803002} {arXiv:hep-th/9803002} \BibitemShut {NoStop}%
\bibitem [{\citenamefont {Rey}\ and\ \citenamefont {Yee}(2001)}]{Rey:1998ik}%
  \BibitemOpen
  \bibfield  {author} {\bibinfo {author} {\bibfnamefont {S.-J.}\ \bibnamefont {Rey}}\ and\ \bibinfo {author} {\bibfnamefont {J.-T.}\ \bibnamefont {Yee}},\ }\href {\doibase 10.1007/s100520100799} {\bibfield  {journal} {\bibinfo  {journal} {Eur. Phys. J. C}\ }\textbf {\bibinfo {volume} {22}},\ \bibinfo {pages} {379} (\bibinfo {year} {2001})},\ \Eprint {http://arxiv.org/abs/hep-th/9803001} {arXiv:hep-th/9803001} \BibitemShut {NoStop}%
\bibitem [{\citenamefont {Drukker}\ \emph {et~al.}(1999)\citenamefont {Drukker}, \citenamefont {Gross},\ and\ \citenamefont {Ooguri}}]{Drukker:1999zq}%
  \BibitemOpen
  \bibfield  {author} {\bibinfo {author} {\bibfnamefont {N.}~\bibnamefont {Drukker}}, \bibinfo {author} {\bibfnamefont {D.~J.}\ \bibnamefont {Gross}}, \ and\ \bibinfo {author} {\bibfnamefont {H.}~\bibnamefont {Ooguri}},\ }\href {\doibase 10.1103/PhysRevD.60.125006} {\bibfield  {journal} {\bibinfo  {journal} {Phys. Rev. D}\ }\textbf {\bibinfo {volume} {60}},\ \bibinfo {pages} {125006} (\bibinfo {year} {1999})},\ \Eprint {http://arxiv.org/abs/hep-th/9904191} {arXiv:hep-th/9904191} \BibitemShut {NoStop}%
\bibitem [{\citenamefont {Louko}\ \emph {et~al.}(2000)\citenamefont {Louko}, \citenamefont {Marolf},\ and\ \citenamefont {Ross}}]{Louko:2000tp}%
  \BibitemOpen
  \bibfield  {author} {\bibinfo {author} {\bibfnamefont {J.}~\bibnamefont {Louko}}, \bibinfo {author} {\bibfnamefont {D.}~\bibnamefont {Marolf}}, \ and\ \bibinfo {author} {\bibfnamefont {S.~F.}\ \bibnamefont {Ross}},\ }\href {\doibase 10.1103/PhysRevD.62.044041} {\bibfield  {journal} {\bibinfo  {journal} {Phys. Rev. D}\ }\textbf {\bibinfo {volume} {62}},\ \bibinfo {pages} {044041} (\bibinfo {year} {2000})},\ \Eprint {http://arxiv.org/abs/hep-th/0002111} {arXiv:hep-th/0002111} \BibitemShut {NoStop}%
\bibitem [{\citenamefont {Giddings}\ and\ \citenamefont {Lippert}(2002)}]{Giddings:2001pt}%
  \BibitemOpen
  \bibfield  {author} {\bibinfo {author} {\bibfnamefont {S.~B.}\ \bibnamefont {Giddings}}\ and\ \bibinfo {author} {\bibfnamefont {M.}~\bibnamefont {Lippert}},\ }\href {\doibase 10.1103/PhysRevD.65.024006} {\bibfield  {journal} {\bibinfo  {journal} {Phys. Rev. D}\ }\textbf {\bibinfo {volume} {65}},\ \bibinfo {pages} {024006} (\bibinfo {year} {2002})},\ \Eprint {http://arxiv.org/abs/hep-th/0103231} {arXiv:hep-th/0103231} \BibitemShut {NoStop}%
\bibitem [{\citenamefont {Czech}\ \emph {et~al.}(2015)\citenamefont {Czech}, \citenamefont {Lamprou}, \citenamefont {McCandlish},\ and\ \citenamefont {Sully}}]{Czech:2015qta}%
  \BibitemOpen
  \bibfield  {author} {\bibinfo {author} {\bibfnamefont {B.}~\bibnamefont {Czech}}, \bibinfo {author} {\bibfnamefont {L.}~\bibnamefont {Lamprou}}, \bibinfo {author} {\bibfnamefont {S.}~\bibnamefont {McCandlish}}, \ and\ \bibinfo {author} {\bibfnamefont {J.}~\bibnamefont {Sully}},\ }\href {\doibase 10.1007/JHEP10(2015)175} {\bibfield  {journal} {\bibinfo  {journal} {JHEP}\ }\textbf {\bibinfo {volume} {10}},\ \bibinfo {pages} {175} (\bibinfo {year} {2015})},\ \Eprint {http://arxiv.org/abs/1505.05515} {arXiv:1505.05515 [hep-th]} \BibitemShut {NoStop}%
\bibitem [{\citenamefont {Cao}\ \emph {et~al.}(2020)\citenamefont {Cao}, \citenamefont {Qi}, \citenamefont {Swingle},\ and\ \citenamefont {Tang}}]{Cao:2020uvb}%
  \BibitemOpen
  \bibfield  {author} {\bibinfo {author} {\bibfnamefont {C.}~\bibnamefont {Cao}}, \bibinfo {author} {\bibfnamefont {X.-L.}\ \bibnamefont {Qi}}, \bibinfo {author} {\bibfnamefont {B.}~\bibnamefont {Swingle}}, \ and\ \bibinfo {author} {\bibfnamefont {E.}~\bibnamefont {Tang}},\ }\href {\doibase 10.1007/JHEP12(2020)033} {\bibfield  {journal} {\bibinfo  {journal} {JHEP}\ }\textbf {\bibinfo {volume} {12}},\ \bibinfo {pages} {033} (\bibinfo {year} {2020})},\ \Eprint {http://arxiv.org/abs/2007.00004} {arXiv:2007.00004 [hep-th]} \BibitemShut {NoStop}%
\bibitem [{\citenamefont {Hamilton}\ \emph {et~al.}(2006{\natexlab{a}})\citenamefont {Hamilton}, \citenamefont {Kabat}, \citenamefont {Lifschytz},\ and\ \citenamefont {Lowe}}]{Hamilton:2005ju}%
  \BibitemOpen
  \bibfield  {author} {\bibinfo {author} {\bibfnamefont {A.}~\bibnamefont {Hamilton}}, \bibinfo {author} {\bibfnamefont {D.~N.}\ \bibnamefont {Kabat}}, \bibinfo {author} {\bibfnamefont {G.}~\bibnamefont {Lifschytz}}, \ and\ \bibinfo {author} {\bibfnamefont {D.~A.}\ \bibnamefont {Lowe}},\ }\href {\doibase 10.1103/PhysRevD.73.086003} {\bibfield  {journal} {\bibinfo  {journal} {Phys. Rev. D}\ }\textbf {\bibinfo {volume} {73}},\ \bibinfo {pages} {086003} (\bibinfo {year} {2006}{\natexlab{a}})},\ \Eprint {http://arxiv.org/abs/hep-th/0506118} {arXiv:hep-th/0506118} \BibitemShut {NoStop}%
\bibitem [{\citenamefont {Hamilton}\ \emph {et~al.}(2006{\natexlab{b}})\citenamefont {Hamilton}, \citenamefont {Kabat}, \citenamefont {Lifschytz},\ and\ \citenamefont {Lowe}}]{Hamilton:2006az}%
  \BibitemOpen
  \bibfield  {author} {\bibinfo {author} {\bibfnamefont {A.}~\bibnamefont {Hamilton}}, \bibinfo {author} {\bibfnamefont {D.~N.}\ \bibnamefont {Kabat}}, \bibinfo {author} {\bibfnamefont {G.}~\bibnamefont {Lifschytz}}, \ and\ \bibinfo {author} {\bibfnamefont {D.~A.}\ \bibnamefont {Lowe}},\ }\href {\doibase 10.1103/PhysRevD.74.066009} {\bibfield  {journal} {\bibinfo  {journal} {Phys. Rev. D}\ }\textbf {\bibinfo {volume} {74}},\ \bibinfo {pages} {066009} (\bibinfo {year} {2006}{\natexlab{b}})},\ \Eprint {http://arxiv.org/abs/hep-th/0606141} {arXiv:hep-th/0606141} \BibitemShut {NoStop}%
\bibitem [{\citenamefont {Almheiri}\ \emph {et~al.}(2019)\citenamefont {Almheiri}, \citenamefont {Engelhardt}, \citenamefont {Marolf},\ and\ \citenamefont {Maxfield}}]{Almheiri:2019psf}%
  \BibitemOpen
  \bibfield  {author} {\bibinfo {author} {\bibfnamefont {A.}~\bibnamefont {Almheiri}}, \bibinfo {author} {\bibfnamefont {N.}~\bibnamefont {Engelhardt}}, \bibinfo {author} {\bibfnamefont {D.}~\bibnamefont {Marolf}}, \ and\ \bibinfo {author} {\bibfnamefont {H.}~\bibnamefont {Maxfield}},\ }\href {\doibase 10.1007/JHEP12(2019)063} {\bibfield  {journal} {\bibinfo  {journal} {JHEP}\ }\textbf {\bibinfo {volume} {12}},\ \bibinfo {pages} {063} (\bibinfo {year} {2019})},\ \Eprint {http://arxiv.org/abs/1905.08762} {arXiv:1905.08762 [hep-th]} \BibitemShut {NoStop}%
\bibitem [{\citenamefont {Penington}(2020)}]{Penington:2019npb}%
  \BibitemOpen
  \bibfield  {author} {\bibinfo {author} {\bibfnamefont {G.}~\bibnamefont {Penington}},\ }\href {\doibase 10.1007/JHEP09(2020)002} {\bibfield  {journal} {\bibinfo  {journal} {JHEP}\ }\textbf {\bibinfo {volume} {09}},\ \bibinfo {pages} {002} (\bibinfo {year} {2020})},\ \Eprint {http://arxiv.org/abs/1905.08255} {arXiv:1905.08255 [hep-th]} \BibitemShut {NoStop}%
\bibitem [{\citenamefont {Brown}\ \emph {et~al.}(2020)\citenamefont {Brown}, \citenamefont {Gharibyan}, \citenamefont {Penington},\ and\ \citenamefont {Susskind}}]{Brown:2019rox}%
  \BibitemOpen
  \bibfield  {author} {\bibinfo {author} {\bibfnamefont {A.~R.}\ \bibnamefont {Brown}}, \bibinfo {author} {\bibfnamefont {H.}~\bibnamefont {Gharibyan}}, \bibinfo {author} {\bibfnamefont {G.}~\bibnamefont {Penington}}, \ and\ \bibinfo {author} {\bibfnamefont {L.}~\bibnamefont {Susskind}},\ }\href {\doibase 10.1007/JHEP08(2020)121} {\bibfield  {journal} {\bibinfo  {journal} {JHEP}\ }\textbf {\bibinfo {volume} {08}},\ \bibinfo {pages} {121} (\bibinfo {year} {2020})},\ \Eprint {http://arxiv.org/abs/1912.00228} {arXiv:1912.00228 [hep-th]} \BibitemShut {NoStop}%
\bibitem [{\citenamefont {Bao}\ \emph {et~al.}(2020{\natexlab{b}})\citenamefont {Bao}, \citenamefont {Chatwin-Davies},\ and\ \citenamefont {Remmen}}]{Bao:2020hsc}%
  \BibitemOpen
  \bibfield  {author} {\bibinfo {author} {\bibfnamefont {N.}~\bibnamefont {Bao}}, \bibinfo {author} {\bibfnamefont {A.}~\bibnamefont {Chatwin-Davies}}, \ and\ \bibinfo {author} {\bibfnamefont {G.~N.}\ \bibnamefont {Remmen}},\ }\href {\doibase 10.1007/JHEP09(2020)102} {\bibfield  {journal} {\bibinfo  {journal} {JHEP}\ }\textbf {\bibinfo {volume} {09}},\ \bibinfo {pages} {102} (\bibinfo {year} {2020}{\natexlab{b}})},\ \Eprint {http://arxiv.org/abs/2006.10762} {arXiv:2006.10762 [hep-th]} \BibitemShut {NoStop}%
\bibitem [{\citenamefont {Heckman}\ \emph {et~al.}(2024)\citenamefont {Heckman}, \citenamefont {H\"ubner},\ and\ \citenamefont {Murdia}}]{Heckman:2024oot}%
  \BibitemOpen
  \bibfield  {author} {\bibinfo {author} {\bibfnamefont {J.~J.}\ \bibnamefont {Heckman}}, \bibinfo {author} {\bibfnamefont {M.}~\bibnamefont {H\"ubner}}, \ and\ \bibinfo {author} {\bibfnamefont {C.}~\bibnamefont {Murdia}},\ }\href {\doibase 10.1103/PhysRevD.110.046007} {\bibfield  {journal} {\bibinfo  {journal} {Phys. Rev. D}\ }\textbf {\bibinfo {volume} {110}},\ \bibinfo {pages} {046007} (\bibinfo {year} {2024})},\ \Eprint {http://arxiv.org/abs/2401.09538} {arXiv:2401.09538 [hep-th]} \BibitemShut {NoStop}%
\bibitem [{\citenamefont {Akers}\ \emph {et~al.}(2024{\natexlab{b}})\citenamefont {Akers}, \citenamefont {Bouland}, \citenamefont {Chen}, \citenamefont {Kohler}, \citenamefont {Metger},\ and\ \citenamefont {Vazirani}}]{Akers:2024wre}%
  \BibitemOpen
  \bibfield  {author} {\bibinfo {author} {\bibfnamefont {C.}~\bibnamefont {Akers}}, \bibinfo {author} {\bibfnamefont {A.}~\bibnamefont {Bouland}}, \bibinfo {author} {\bibfnamefont {L.}~\bibnamefont {Chen}}, \bibinfo {author} {\bibfnamefont {T.}~\bibnamefont {Kohler}}, \bibinfo {author} {\bibfnamefont {T.}~\bibnamefont {Metger}}, \ and\ \bibinfo {author} {\bibfnamefont {U.}~\bibnamefont {Vazirani}},\ }\href@noop {} {\  (\bibinfo {year} {2024}{\natexlab{b}})},\ \Eprint {http://arxiv.org/abs/2411.04978} {arXiv:2411.04978 [hep-th]} \BibitemShut {NoStop}%
\bibitem [{\citenamefont {Bhattacharjee}\ and\ \citenamefont {Naskar}(2024)}]{Bhattacharjee:2024ceb}%
  \BibitemOpen
  \bibfield  {author} {\bibinfo {author} {\bibfnamefont {A.}~\bibnamefont {Bhattacharjee}}\ and\ \bibinfo {author} {\bibfnamefont {J.}~\bibnamefont {Naskar}},\ }\href@noop {} {\  (\bibinfo {year} {2024})},\ \Eprint {http://arxiv.org/abs/2411.02825} {arXiv:2411.02825 [hep-th]} \BibitemShut {NoStop}%
\bibitem [{\citenamefont {Pastawski}\ and\ \citenamefont {Preskill}(2017)}]{Pastawski:2016qrs}%
  \BibitemOpen
  \bibfield  {author} {\bibinfo {author} {\bibfnamefont {F.}~\bibnamefont {Pastawski}}\ and\ \bibinfo {author} {\bibfnamefont {J.}~\bibnamefont {Preskill}},\ }\href {\doibase 10.1103/PhysRevX.7.021022} {\bibfield  {journal} {\bibinfo  {journal} {Phys. Rev. X}\ }\textbf {\bibinfo {volume} {7}},\ \bibinfo {pages} {021022} (\bibinfo {year} {2017})},\ \Eprint {http://arxiv.org/abs/1612.00017} {arXiv:1612.00017 [quant-ph]} \BibitemShut {NoStop}%
\bibitem [{\citenamefont {Bao}\ and\ \citenamefont {Naskar}(2023)}]{Bao:2022tgv}%
  \BibitemOpen
  \bibfield  {author} {\bibinfo {author} {\bibfnamefont {N.}~\bibnamefont {Bao}}\ and\ \bibinfo {author} {\bibfnamefont {J.}~\bibnamefont {Naskar}},\ }\href {\doibase 10.1103/PhysRevD.107.066014} {\bibfield  {journal} {\bibinfo  {journal} {Phys. Rev. D}\ }\textbf {\bibinfo {volume} {107}},\ \bibinfo {pages} {066014} (\bibinfo {year} {2023})},\ \Eprint {http://arxiv.org/abs/2209.15026} {arXiv:2209.15026 [hep-th]} \BibitemShut {NoStop}%
\end{thebibliography}%

\end{document}